\documentclass[12pt,preprint]{aastex}

\def\msun{{\rm\,M_\odot}}
\newcommand{\etal}{et al.\ }

\newcommand{\lya}{Ly$\alpha$ }

\begin{document}

\title{A Constraint on the Gravitational Lensing Magnification and Age
of the Redshift z=6.28 Quasar SDSS 1030+0524}

\author{Zolt\'an Haiman\altaffilmark{1} and Renyue Cen}
\affil{Princeton University Observatory, Princeton, NJ 08544, USA; 
zoltan,cen@astro.princeton.edu}
\altaffiltext{1} {Hubble Fellow}

\accepted{ }

\begin{abstract}

The recent discovery of bright quasars at redshift $z\sim 6$ suggests that
black holes (BHs) with masses in excess of $\sim10^9~{\rm M_\odot}$ have
already assembled at a very early stage in the evolution of the universe.  An
alternative interpretation is that these quasars are powered by less massive
BHs, but their fluxes are strongly magnified through gravitational lensing by
intervening galaxies.  Here we analyze the flux distribution of the \lya
emission of the quasar with the highest known redshift, SDSS 1030+0524, at
$z=6.28$.  We show that this object could not have been magnified by lensing by
more than a factor of $\sim 5$.  The constraint arises from the large observed
size, $\sim$30 (comoving) Mpc, of the ionized region around this quasar, and
relies crucially only on the assumption that the quasar is embedded in a
largely neutral IGM.  Based on the line/continuum ratio of SDSS 1030+0524, we
argue further that this quasar also cannot be beamed by a significant
factor. We conclude that the minimum mass for its resident BH is $4\times
10^8~{\rm M_\odot}$ (for magnification by a factor of 5); if the mass is this
low, then the quasars had to switch on prior to redshift $z_f\ga9$.  From the
size of the ionized region, we are also able to place an absolute lower bound
on the age of this quasar at $t>2\times 10^{7}~$yrs.

\end{abstract}


\section{Introduction}

Supermassive black holes (SMBH) are believed to power quasars and active
galactic nuclei (AGN, see, e.g., Rees 1984, 1990) and extragalactic jets
(Blandford 1989).  There is mounting evidence that even normal (i.e.,
non-active) galaxies, including our own, harbor SMBHs in their centers, with
black hole masses in the range of $10^6-10^9\msun$ (Magorrian \etal 1998;
Richstone \etal 1998; Genzel \& Eckart 1999; Tremaine \etal 2002).  There are
many unanswered fundamental questions concerning how and when SMBHs formed and
grew (for an exhaustive list of SMBH formation mechanisms the reader is
referred to Rees 1984).  However, there is now compelling evidence that bright
quasar activity reflects the growth of SMBHs by accretion (e.g. Lynden-Bell
1967; Soltan 1982).

The total light output of known quasars, summed over all redshifts,
corresponds to $\epsilon\approx 10\%$ of the rest--mass energy of the
total BH mass in their presumed remnants at redshift $z=0$ (see, e.g.,
the recent paper Yu \& Tremaine 2002, and references therein).
Conversely, this implies that SMBHs have grown most of their mass in a
luminous phase, accreting no faster than the Eddington rate,
corresponding to exponential growth on the timescale of $t_{\rm
Edd}=4\times 10^7~(\epsilon/0.1)~{\rm yr}$.  With the recent discovery
of four quasars around redshift $z\approx 6$ in the Sloan Digital Sky
Survey (SDSS), it appears that bright quasar activity already
commenced at this early epoch.  Under the assumption that these quasar
BHs shine at their Eddington limit, and they are not gravitationally
lensed, or beamed, their masses are $\sim (2-3)\times 10^9\msun$.  As
emphasized by Haiman \& Loeb (2001), the time it takes for a SMBH to
grow to this size from a stellar seed of $\sim 100\msun$ is $4\times
10^7~{\rm ln}(3\times10^9/100)~{\rm yr}\sim 7\times 10^8~{\rm yr}$,
comparable to the age of the universe at $z=6$ ($\sim 9\times
10^8~{\rm yr}$ for a flat $\Lambda$CDM universe with $H_0=70~{\rm
km/s/Mpc}$ and $\Omega_m=0.3$).

SMBHs at such an early stage in the evolution of the universe can thus be only
marginally accommodated into our currently popular cosmological
models. Furthermore, their existence implies that the seeds of these SMBHs had
to appear at very high redshift ($z\ga 10$), and also places constraints on the
physics of how SMBHs grow. In particular, their growth could not have been much
slower than the Eddington rate under a nominal $10\%$ radiative efficiency, and
high efficiencies ($\ga 20\%$) can already be ruled out (see, e.g., Haiman \&
Loeb 2001 for details).

Such conclusions rely on the fact that the SMBHs are indeed very massive.  The
masses of SMBHs in luminous quasars are uncertain to the extent that the
quasars' emission may be magnified through gravitational lensing by foreground
objects (e.g. Turner 1991).  Indeed, it is possible to consider the statistics
of intervening lenses, together with a luminosity function (LF) of $z\sim 6$
quasars, to estimate the expected probability that the SDSS quasars were
significantly magnified. The $z=6$ quasar LF is very poorly constrained, and
even strong magnifications can have large probabilities if the LF is steep
(Comerford et al. 2002; Wyithe \& Loeb 2002).  Strong magnification of the SDSS
quasars could therefore constitute an alternative explanation to their high BH
masses. To a lesser level, the inferred masses of the SMBHs are also subject to
variations of parameters in the assumed cosmological model.

A separate question concerns the lifetime of quasars. Unlike the lifetime of
stars, which may be computed accurately theoretically, the lifetime of quasars
is largely unknown due to large uncertainties in black hole accretion physics
(i.e., radiation efficiency) and gas supply.  As a result, constraints on the
lifetime of quasars can only be placed empirically and indirectly.  The very
convincing evidence for the existence of massive black holes in majority of
normal galaxies (Magorrian \etal 1998; Richstone \etal 1998; Genzel \& Eckart
1999; Tremaine \etal 2002) implies that a typical lifetime of quasars of $\sim
t_{Hubble} n_Q/n_G \sim 10^7~$yrs, where $t_{Hubble}$ is the Hubble time, $n_Q$
and $n_G$ the observed number density of quasars and normal galaxies,
respectively (Blandford 1999).  An upper limit on quasar lifetime of $10^9~$yrs
may be inferred based on the low redshift ($z<3$) decrease of quasar abundance.
The comparison of quasar clustering properties and simulated galaxy halos of
comparable clustering properties can also be used to infer a quasar lifetime of
$10^6-10^8~$yrs (Haiman \& Hui 2001; Martini \& Weinberg 2001).  Finally, a
quasar lifetime of $\sim 10^7~$yrs has recently been inferred from the size of
the HeIII proximity region of a quasar embedded in an IGM where helium is
mostly in the form of HeII (at $z\ga 3$, see, e.g., Anderson et al. 1999; this
is similar to our own method).

In this {\it Letter}, we focus our attention on the bright quasar with the
highest known redshift, SDSS 1030+0524 at $z=6.28$. We analyze the spectrum of
this source to derive interesting constraints both on the BH mass, and on the
age of this source. This source is unique among the four $z\sim 6$ quasars, in
that it shows a strong Gunn-Peterson trough, and appears to be embedded in a
neutral intergalactic medium (IGM).  The spectrum of a point source of \lya
radiation would show no flux shortward of the central \lya wavelength, due to
absorption by the intervening neutral IGM.  However, some flux may be
transmitted shortward of the \lya line, provided that the source is surrounded
by a sufficiently large local ionized region (Str\"omgren sphere).  In an
earlier paper (Cen \& Haiman 2000; hereafter CH) we showed that the
transmission of the blue side of the intrinsic \lya emission line directly
depends on the total number of ionizing photons emitted over the lifetime of
the source: i.e. on the product of its lifetime and its intrinsic luminosity
(or BH mass).  This situation is unique for quasars embedded in a neutral IGM.
For less luminous objects, such as galaxies, the damping wing of the resonance
\lya absorption of the intervening neutral IGM (Miralda-Escud\'e 1998) would
cast a very high optical depth and render most of the intrinsic \lya emission
invisible.  Likewise, a quasar embedded in a highly ionized IGM at lower
redshifts rapidly establishes ionization equilibrium, and the 'proximity
effect' in its spectrum (Bajtlik, Duncan, \& Ostriker 1988) reveals no
constraint about its age.  However, a bright quasar at high redshift, prior to
cosmological reionization, such as SDSS 1030+0524, can provide a unique way to
tightly constrain both the black hole mass and the age of the quasar.  In this
{\it Letter} we analyze the spectrum of SDSS 1030+0524 to provide such a
quantitative constraint.  

The rest of this paper is organized as follows. In \S~2, we describe our model
for the spectrum of the Ly$\alpha$ line of a high redshift source. In \S~3, we
apply this model to SDSS 1030+0524 to derive a constraint on its lensing
magnification and its lifetime. In \S~4, we discuss the implications of our
results and their caveats.  Finally, in \S~5, we offer our
conclusions. Throughout this paper, we assume the background cosmology to be
flat $\Lambda$CDM with $(\Omega_\Lambda,\Omega_{\rm m},\Omega_{\rm
b},h)=(0.7,0.3,0.04,0.7)$.

\section{Transmission of the \lya Emission Line}

In this section, we simulate the observed spectrum of a quasar near its \lya
emission line, for a source embedded in a neutral IGM, with a given age and BH
mass.  Our modeling has two basic steps: (i) we place a quasar in the neutral
IGM, and compute the size of its ionized (HII) region, and (ii) we simulate the
spectrum of the emission line by assuming an intrinsic template lineshape, and
including the effects of the absorbing gas both within and outside the HII
region.  For a detailed description of the method, the reader is referred to
CH. Here we briefly recapitulate.

Ignoring any recombinations, the radius $R_{\rm s}$ of the Str\"omgren sphere of a
quasar of age $t_Q$ is such that the number of hydrogen atoms in the sphere
equals the total number of ionizing photons produced by the source:

\begin{eqnarray}
\label{eq:Rt1}
 R_{\rm s} =  \left[ \frac{3\dot N_{\rm ph} t_Q}{4\pi \langle n_H \rangle} 
      \right]^{1/3}
    & = & 4.5~\left(\frac{N_{\rm ph}}{1.3\times 10^{57}~{\rm s^{-1}}}
          \right)^{1/3}
\left(\frac{t_Q}{2\times10^7~{\rm yr}}\right)^{1/3}
\left(\frac{1+z_Q}{7.28}\right)^{-1}
\,     {\rm Mpc}\\
    & = & 4.5~\left(\frac{M_{\rm bh}}{2\times 10^{9}~{\rm M_\odot}}
          \right)^{1/3}
\left(\frac{t_Q}{2\times10^7~{\rm yr}}\right)^{1/3}
\left(\frac{1+z_Q}{7.28}\right)^{-1}
\,     {\rm Mpc}
\label{eq:Rt2}
\end{eqnarray}

\noindent 
(CH), where $R_{\rm s}$ is in proper (not comoving) units, $\langle n_H \rangle$ is
the mean hydrogen density within $R_{\rm s}$, and $\dot N_{\rm ph}$ is emission rate
of ionizing photons from the quasar.  For $\langle n_H \rangle$, we have
adopted the mean IGM density at $z=6.28$ with $\Omega_{\rm b}=0.04$, and $\dot
N_{\rm ph}=1.3\times 10^{57}~{\rm s^{-1}}$ is calibrated from the observed flux
of SDSS 1030+0524, extrapolating its continuum to energies above 13.6eV using a
standard quasar template spectrum (Elvis et al. 1994, which agrees well with
the more recent SDSS template in Vanden Berk et al. 2001, hereafter VdB).
Adopting the same template spectrum, and assuming further that the BH powering
this quasar is shining at its Eddington luminosity, the mass of the BH is found
to be $M_{\rm bh}=2\times10^9~{\rm M_\odot}$. We have used this mass to convert
equation~(\ref{eq:Rt1}) into equation~(\ref{eq:Rt2}).

In an evolving and clumpy cosmological density field around a quasar, ionizing
photons are lost to recombinations, and the size of the HII region can be
reduced relative to the prediction of equation~(\ref{eq:Rt2}). We need to then
solve the equation for the radius of the ionization front, $R_{\rm s}$, taking into
account all relevant effects including recombination, density evolution and
cosmological effects as follows:

\begin{equation}
\frac{dR_{\rm s}^3}{dt}= 3H(z) R_{\rm s}^3 + \frac{3 \dot N_{\rm ph}}{4\pi \langle n_H
 \rangle} - C \langle n_H \rangle \alpha_B R_{\rm s}^3,
\label{eq:Ri}
\end{equation}

\noindent
where $H(z)$ is the Hubble constant at $z$, $C\equiv\langle n_H^2 \rangle /
\langle n_H \rangle^2$ is the mean clumping factor of ionized gas within $R_{\rm s}$,
and $\alpha_B$ is the hydrogen recombination coefficient.  The three terms on
the right side of equation~(\ref{eq:Ri}) account for the Hubble expansion, the
ionizations by newly produced photons, and recombinations, respectively
(Shapiro \& Giroux 1987; Haiman \& Loeb 1997).  Although
equation~(\ref{eq:Rt2}) is accurate for low clumping factors and quasar ages
($C\la 10$, $t_Q\la 10^8$ yrs), the results presented here are based on a
numerical solution of equation~(\ref{eq:Ri}).

In general, a quasar has an intrinsic Ly$\alpha$ emission line, which is
reprocessed along the line of sight to the observer by the opacity of the
intervening neutral IGM as well as of the residual neutral hydrogen within the
Str\"omgren sphere.  We have to here model the absorption in both regions. The
IGM outside the HII region is assumed to be neutral (see discussion
below). Inside the HII region, we model the density distribution using the
hydrodynamical simulation described in Cen \& McDonald (2002).  For
illustrative purposes, we have chosen here to focus on a single line of sight
through this simulation box, which has a mean gas clumping factor of $C=10$,
and an approximate log--normal density distribution. A full statistical
description of the expected flux distribution within the HII region, based on a
large sample of lines of sights, is not crucial for our present purposes, but
will be considered in a future paper (Haiman \& Cen, in preparation).

Finally, in order to model the intrinsic profile of the emission line, we
utilize the median \lya emission line shape as observationally determined by
VdB.  We first obtain the line profile from the VdB spectrum by subtracting a
constant $6.3\times10^{-17}~{\rm~erg~s^{-1}~cm^{-2}}$ \AA$^{-1}$, which is
approximately the median observed continuum level.  To fit the red (unabsorbed)
side of the Ly$\alpha$ line of SDSS 1030+0524, we find we have to divide the
VdB profile by a factor of $\approx 3$. To approximate the observed continuum
of SDSS 1030+0524, we add back a constant of
$1.1\times10^{-17}~{\rm~erg~s^{-1}~cm^{-2}}$ \AA$^{-1}$.  In other words, the
observed continuum flux of SDSS 1030+0524 is about 6 times fainter than that of
the median $z>2.25$ quasar, while the emission line is only about 3 times
fainter.  The VdB spectrum is the median spectrum of about $\sim$150 quasars at
$z>2.25$, with a mean redshift of $z\approx 3$.  The IGM has an average opacity
of $\tau_{\rm IGM}\sim 0.4$ at the blue side of the Ly$\alpha$ line at this
redshift (e.g. Madau 1995). To create our intrinsic template line shape between
1200 and 1215\AA\, (8750 and 8850\AA\, observed), we divide the VdB profile by
a correction factor that varies linearly from unity at 1215\AA\, to
$\exp(-\tau_{\rm IGM})$ at 1200\AA.  This is intended to take into account the
fact that each quasar in the VdB sample suffers from both IGM opacity and from
its own proximity effect (so that there is no correction at the line center).

\begin{figure}
\plotone{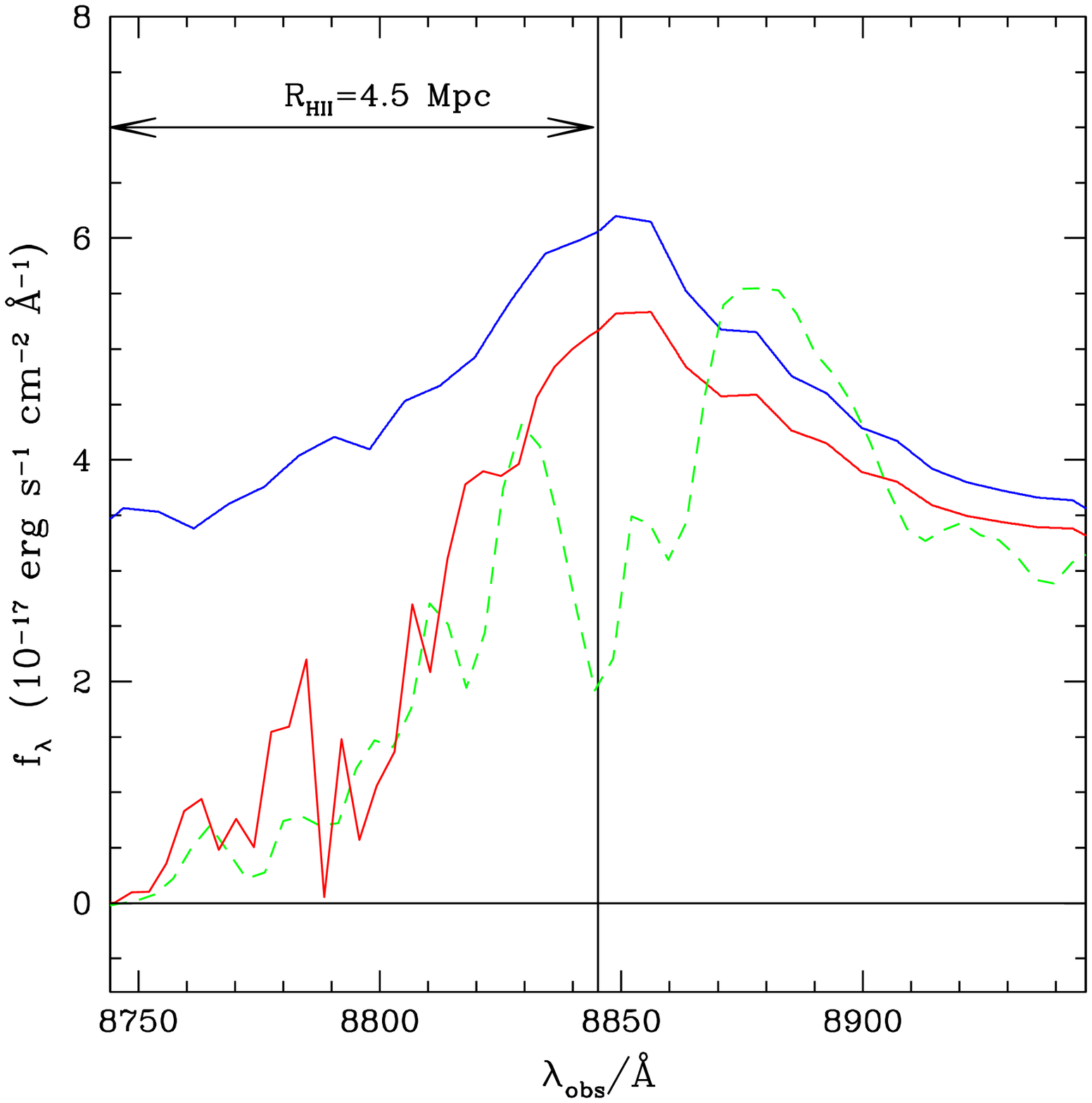}
\caption{The spectrum of SDSS 1030+0524 in the vicinity of its Ly$\alpha$
emission line.  The dashed curve indicates the observed spectrum from Becker et
al. (2002).  The solid curves show the expected line profile based on the mean
spectrum of a sample of $z>2.25$ quasars, without any absorption in the IGM
(i.e. the assumed intrinsic line profile, top curve), as well as the profile
after it is processed through the IGM (bottom curve). All three spectra are
smoothed on a scale of $\sim4$\AA.  The presence of flux on the blue side of
the line down to a wavelength of $8750$\AA\, implies that the source is
surrounded by a large ionized region, of proper size 4.5 Mpc (or 32 comoving
Mpc).}
\label{spec}
\end{figure}

Figure~\ref{spec} shows the resulting intrinsic \lya emission line (top solid
curve), as well as an illustrative example of the processed profile (bottom
solid curve). In this example, we assume a quasar age of $t_Q=2\times 10^{7}$
years, and we adopt the ionizing photon production rate of $\dot N_{\rm
ph}=1.3\times 10^{57}~{\rm s^{-1}}$, corresponding to the observed flux of SDSS
1030+0524 with no lensing corrections.  This results in a proper HII-radius of
4.5 Mpc.  The bottom solid curve in Figure~\ref{spec} shows the resulting
reprocessed \lya emission line, including the HI opacities from both within and
outside the HII region. Also shown in this figure, as the dashed curve, is the
observed flux distribution (Becker \etal 2001).

Our final processed spectrum is in reasonable agreement with the data.  There
is a discrepancy at the \lya line center: our model does not account for the
large apparent additional opacity centered at the quasar redshift.  However,
this does not effect our conclusions drawn from the blue tail of the \lya line.
The crucial feature of the observed spectrum of SDSS 1030+0524 is the presence
of significant flux down to 8750\AA.  Any model that allows for the presence of
this flux must have an HII region whose size is at least $>4.5$ Mpc, as is the
case in the example shown in Figure~\ref{spec}. If the HII region was smaller,
the flux at wavelengths above 8750\AA\, would be suppressed by an enormous
factor [$\sim \exp(10^6)$] from the Gunn-Peterson opacity of the neutral IGM.

\section{Constraints on Lensing Magnification and the Age of SDSS 1030+0524}

As we have seen above, the presence of flux in the spectrum of SDSS 1030+0524
down to 8750\AA\, implies that this quasar is surrounded by a large (4.5 proper
Mpc) Str\"omgren sphere.  Assuming that its apparent flux reflects its true
luminosity, this source is powered by a black hole whose mass is $M_{\rm
bh}=2\times10^9~{\rm M_\odot}$, and, from equation~(\ref{eq:Rt1}), its age has
to be at least $2\times 10^7$ years.\footnote{The age could be shorter(longer)
if the source was brighter(fainter) in the past.  However, SDSS 1030+0524 is
already unusually bright; being even brighter would make the BH even more
puzzlingly massive.}

However, the flux of SDSS 1030+0524 may be magnified through gravitational
lensing by foreground galaxies (or stars).  As mentioned in the introduction,
the probability for significant magnification can be appreciable, depending on
the shape of the unknown quasar luminosity function (Comerford et al. 2002;
Wyithe \& Loeb 2002).  Obviously, for a given observed flux from the quasar, a
larger gravitational lensing magnification would imply an intrinsically fainter
quasar.  A fainter quasar would, in turn, require a longer age to produce a
Str\"omgren sphere with the required size of at least 4.5 Mpc.  Thus, the
minimum quasar age is an increasing function of gravitational lensing
magnification. Indeed, a constraint on $t_{\rm min}$ may be most informatively
placed in the ($t_{\rm min}$ $vs$ magnification) plane.

\begin{figure}
\plotone{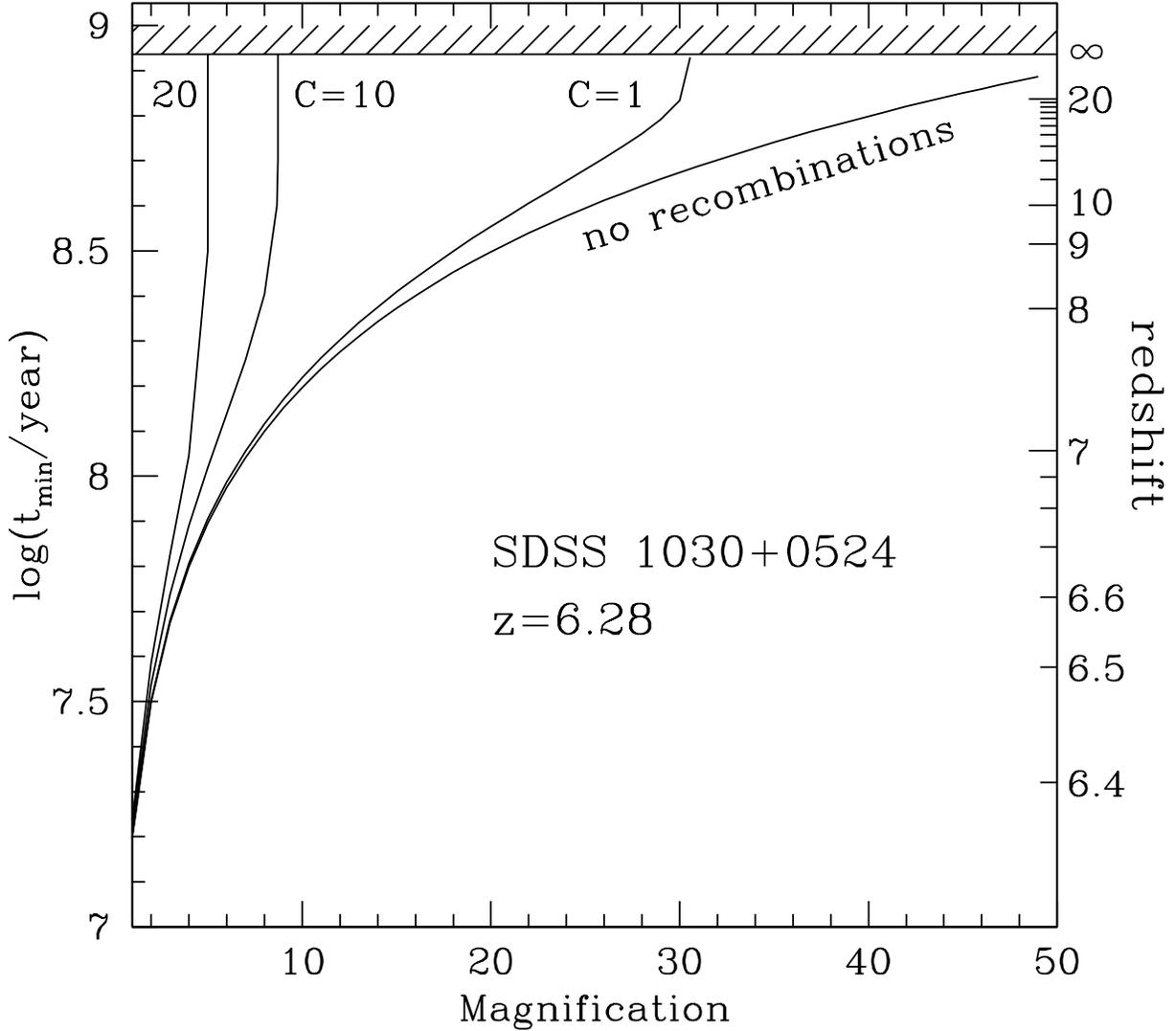}
\caption{The minimum age of SDSS 1030+0524, as a function of its assumed
gravitational lensing magnification. The constraint is based on the requirement
that the radius of the Str\"omgren sphere, computed from
equation~(\ref{eq:Ri}), should be $R_{\rm s}=4.5~$Mpc.  The four curves show
four cases with $\alpha=0$ (ignoring recombinations, in this case $t_{\rm min}$
scales linearly with the magnification), and with three different gas clumping
factors, $C=1$, $C=10$ and $C=20$.  The upper shaded region is excluded,
because it exceeds the age of the universe at $z=6.28$.  For realistic clumping
factors ($C>20$), the magnification cannot exceed a factor of 5 for any age of
the source.}
\label{tmin}
\end{figure}

Figure~\ref{tmin} shows the constraints on the minimum age of SDSS 1030+0524 as
a function of its gravitational lensing magnification. The constraint is based
on the requirement that the radius of the Str\"omgren sphere, computed from
equation~(\ref{eq:Ri}), should be $R_{\rm s}=4.5~$Mpc.  The case presented here
assumes that the IGM is still largely neutral at $z=6.28$.  The four curves
show four cases with $\alpha=0$ (ignoring recombinations, in this case $t_{\rm
min}$ scales linearly with the magnification), and with three different gas
clumping factors, $C=1$, $C=10$ and $C=20$.  The upper shaded region is
excluded, because it exceeds the age of the universe at $z=6.28$.

Several important points may be learned from Figure~\ref{tmin}.  First, $t_{\rm
min}< 2\times 10^{7}~$yrs is not allowed for any value of magnification;
$t_{\rm min}=2\times 10^{7}~$yrs is an {\it absolute lower} limit.  Pentericci
\etal (2001) derive the age of SDSS 1030+0524 to be $1.33\times 10^7~$yrs by
simply counting ionizing photons at $z=6.28$.  They do not take into account
the evolution of the density field and hydrogen recombination.  While we have
performed more detailed calculations, we stress that in either case the derived
age of the quasar can only be a lower bound, since locally (e.g., at
$\lambda\ge 8750$\AA) the transmission of rest-frame \lya photons is determined
by the balance between quasar ionizing flux and hydrogen recombination, once
the required age is reached.

Second, the maximum allowed gravitational lensing magnification is a strong
function of the assumed clumping factor.  For large clumping factors, the size
of the HII region lags increasingly behind its value in the no-recombination
case, and increasingly longer ages are required. Note that for $C>10$, the HII
sphere reaches its (steady-state) equilibrium value in a few $\times 10^8$
years, after which it ceases to grow.  Figure~\ref{tmin} also reveals that for
realistic clumping factors ($C>20$, see discussion below), the magnification
{\it cannot} exceed a factor of $5$.  Even for a clumping factor as low as
$C=10$, the magnification must be less than a factor of 10.  Note that for
$C=10$, if the quasar was magnified by the maximum possible factor of $9$,
$R_{\rm s}=4.5~$Mpc (proper) corresponds to the equilibrium sphere radius,
which is reached in $10^{8.5}~$yrs.  In this case the size of the Str\"omgren
sphere ceases to increase, once its age reaches $10^{8.5}$yrs, regardless when
it was turned on. In other words, if the gravitational lensing magnification is
$9$, the quasar had to be turned on at redshift some redshift prior to $z\sim
10$.  In the more likely case of $C=20$, by the same argument, the quasar could
only have been magnified by a factor of $5$ if it turned on at $z\ga 9$.  To
summarize: each SDSS quasar at $z\sim 6$ could have a significant probability
of having being magnified through gravitational lensing by a factor of $\ge 10$
(Comerford et al. 2002; Wyithe \& Loeb 2002), potentially casting doubt on the
validity of the derived large BH masses.  However, here we find that SDSS
1030+0524 cannot be gravitationally magnified by more than a factor of $\sim
5$.

\section{Discussion}

Our conclusions above, most importantly that SDSS 1030+0524 cannot be an
intrinsically faint and strongly lensed source, relies only on the need for
this source to create a Str\"omgren sphere with a radius at least 4.5 Mpc.  In
particular, the conclusions are insensitive to the detailed modeling of the
density distribution near the quasar. Our most important assumption is that the
IGM is largely neutral at $z=6.28$; a statement supported by the detection of
the Gunn-Peterson trough in the spectrum of this source.  Although it would be
possible to block out the flux of SDSS 1030+0524 at wavelengths shorter than
8750\AA\, even if the neutral fraction was $x_{\rm H}\sim 10^{-2}$ (Becker
\etal 2001; Barkana 2001; Fan \etal 2002), the inferred rapid rise of the
metagalactic radiation field from $z\sim 5.5$ to $z\sim 6.3$ suggests that the
universe is indeed neutral at the redshift of this source (Cen \& McDonald
2002; Gnedin 2002).  A comparison between the observed ionizing background flux
evolution and numerical simulations indeed shows a strong case.  Simulations
have shown that the reionization phase begins with a relatively slow process on
a time scale of about a Hubble time, during which the mean radiation field
builds up to a value of approximately $10^{-24}~$erg/cm$^2$/Hz/sec/sr at the
Lyman limit. This is followed by a brief ``overlap'' phase, when the {\it
majority} of the baryons are ionized, accompanied by a sudden jump in the
amplitude of the mean radiation field intensity at the Lyman limit to
$10^{-22}-10^{-21}~$erg/cm$^2$/Hz/sec/sr, which occurs within a redshift
interval of a fraction of unity (Miralda-Escud\'e \etal 2000; Gnedin 2000).  If
we identify the observed sudden rise of the ionizing radiation background at
$z\sim 6.1$ with the epoch of rapid increase in the ionizing radiation
background seen simulations, the baryons must largely be neutral at $z\sim
6.28$.

An important input to our model is the mean clumping of the gas in the IGM.
Cosmological simulations of the canonical $\Lambda$CDM model yield $C_{\rm
HII}\sim 40$ (Figure 2 in Gnedin \& Ostriker 1997) at $z\sim 6$.  Note that the
model adopted in Gnedin \& Ostriker (1997) has $\sigma_8=0.67$, which is
perhaps rather conservative and consistent with the newer, lower normalization
based on X-ray clusters (Seljak 2001).  A somewhat different clumping factor
may be predicted in a model that self-consistently reionizes the universe at
$z\sim 6$ instead of $z\sim 7$ in the simulation.  However, it seems unlikely
that $C$ can be as low as $10$ for a sensible model.

An alternative way to avoid a large BH mass in SDSS 1030+0524 would be if this
source has a low total luminosity, but is strongly beamed towards us.  A test
of this hypothesis is the line/continuum ratio.  For a given observed continuum
(ionizing) flux, a presence of beaming into a solid angle $\Delta\Omega$ would
reduce the strength of any isotropic emission (recombination) line by a factor
of $\Delta\Omega/4\pi$, since the lines would only be produced only within the
cone into which the ionizing radiation is beamed.  As we have seen above, the
line/continuum ratio of SDSS 1030+0524 is about twice that of the median
$z>2.25$ quasar.  This indicates that SDSS 1030+0524 is less likely to be
significantly beamed than a typical high--redshift optical quasar.  On the
other hand, a typical quasar is indeed unlikely to be significantly beamed: (1)
typical optical quasars do not show relativistic spectral features similar to
those found in BL Lac objects, which are {\it known} to be beamed, (2) there is
no indication why a typical quasar is different from lower luminosity AGNs and
Seyfert galaxies that are {\it known not} to be beamed, (3) there is no natural
mechanism to produce a strongly beamed radiation whose spectrum remains close
to a black--body, as is seen in the ``blue--bump'' component in the spectra of
many quasars, and (4) assuming no beaming, the characteristic luminosities and
abundances of bright quasars near $z\sim 2.5$ are consistent with the
characteristic masses and abundances of their remnant BHs at $z=0$ (see, e.g. Yu
\& Tremaine 2002); strong beaming would likely invalidate this successful
agreement.

Combining this with the finding above that the quasar is not magnified through
gravitational lensing by a significant factor implies that SDSS 1030+0524 needs
to indeed contain a very massive black hole.  Assuming the quasar radiates at
the Eddington luminosity and magnified through gravitational lensing by a
factor of $5$, the minimum mass for its resident BH is $4\times10^8~{\rm
M_\odot}$, and it then has to have formed at redshift $z>9$.

\section{Conclusions}

We have analyzed the flux distribution of the \lya emission of the quasar with
the highest known redshift, SDSS 1030+0524 at $z=6.28$, discovered by the Sloan
Digital Sky Survey.  From its spectrum, we infer the presence of a large ($\sim
4.5$Mpc) ionized region around this QSO.  The large size of this ionized region
makes it impossible for this source to be intrinsically faint, or to be very
young.  We find that SDSS 1030+0524 could not have been magnified through
gravitational lensing by more than a factor of $\sim5$.  The line/continuum
ratio of SDSS 1030+0524 is observed to be twice that of the median $z>2.25$
quasar, indicating that this quasar is also unlikely to be significantly
beamed.  Combining these two facts {\it requires that the minimum mass for its
resident BH is indeed $\sim10^9~M_\odot$}.  If the quasar is not lensed, and is
shining at (or below) the Eddington luminosity of its resident SMBH, then the
inferred mass is at least $M_{\rm bh}=2\times10^9~{\rm M_\odot}$, and its
minimum age is $2\times 10^{7}~$yrs.  These numbers can only be modified by
gravitational lensing by relatively small factors: if the source is magnified
by the maximum allowed factor of $5$, the BH mass is $\sim 4\times10^8~{\rm
M_\odot}$, and in this case, its age has to be longer than $10^8~$yrs, placing
its formation redshift at $z_f>9$.

As the formation of such massive black holes in the universe at such high
redshifts is already presenting a theoretical challenge, it is important to
have limits on the magnification of their fluxes by gravitational lensing.
Although SDSS 1030+0524 is currently the only high redshift quasar to which our
method is applicable, the constraints we have derived for this source can be
repeated and applied to future quasars that will be discovered at $z\ga 6.3$,
prior to the reionization epoch.

\acknowledgments 
We thank Michael Strauss and Dan Vanden Berk for providing the spectrum of SDSS
1030+0524, and the mean SDSS quasar spectrum, in electronic form, together with
helpful narratives.  ZH acknowledges support by NASA through the Hubble
Fellowship grant HF-01119.01-99A, awarded by the Space Telescope Science
Institute, which is operated by the Association of Universities for Research in
Astronomy, Inc., for NASA under contract NAS 5-26555.  RC was supported in part
by grants AST93-18185 and ASC97-40300.

\end{document}